\documentclass{aa}

\usepackage{graphicx}
\usepackage{natbib}
\usepackage{txfonts}
\usepackage{amsmath}
\usepackage[T1]{fontenc}
\usepackage{subcaption}
\usepackage{placeins}
\usepackage[colorlinks=true,allcolors=blue]{hyperref}

\newcommand{\fastspec}{\textsc{FastSpec}}
\newcommand{\atlas}{\textsc{Atlas}}

\newcommand{\synthe}{\textsc{Synthe}}
\newcommand{\sme}{\textsc{SME}}
\newcommand{\pysme}{\textsc{PySME}}
\newcommand{\turbospectrum}{\textsc{Turbospectrum}}
\newcommand{\vald}{\textsc{VALD}}
\newcommand{\exomol}{\textsc{ExoMol}}
\newcommand{\hbop}{\textsc{Hbop}}
\newcommand{\kms}{km\,s$^{-1}$}
\newcommand{\Teff}{T_{\mathrm{eff}}}
\newcommand{\logg}{\log g}
\newcommand{\vsini}{\varv\sin i}
\newcommand{\vmic}{\varv_{\mathrm{mic}}}
\newcommand{\vmac}{\varv_{\mathrm{mac}}}

\newcommand{\orcidlink}[1]{\protect\href{https://orcid.org/#1}{\protect\includegraphics[width=8pt]{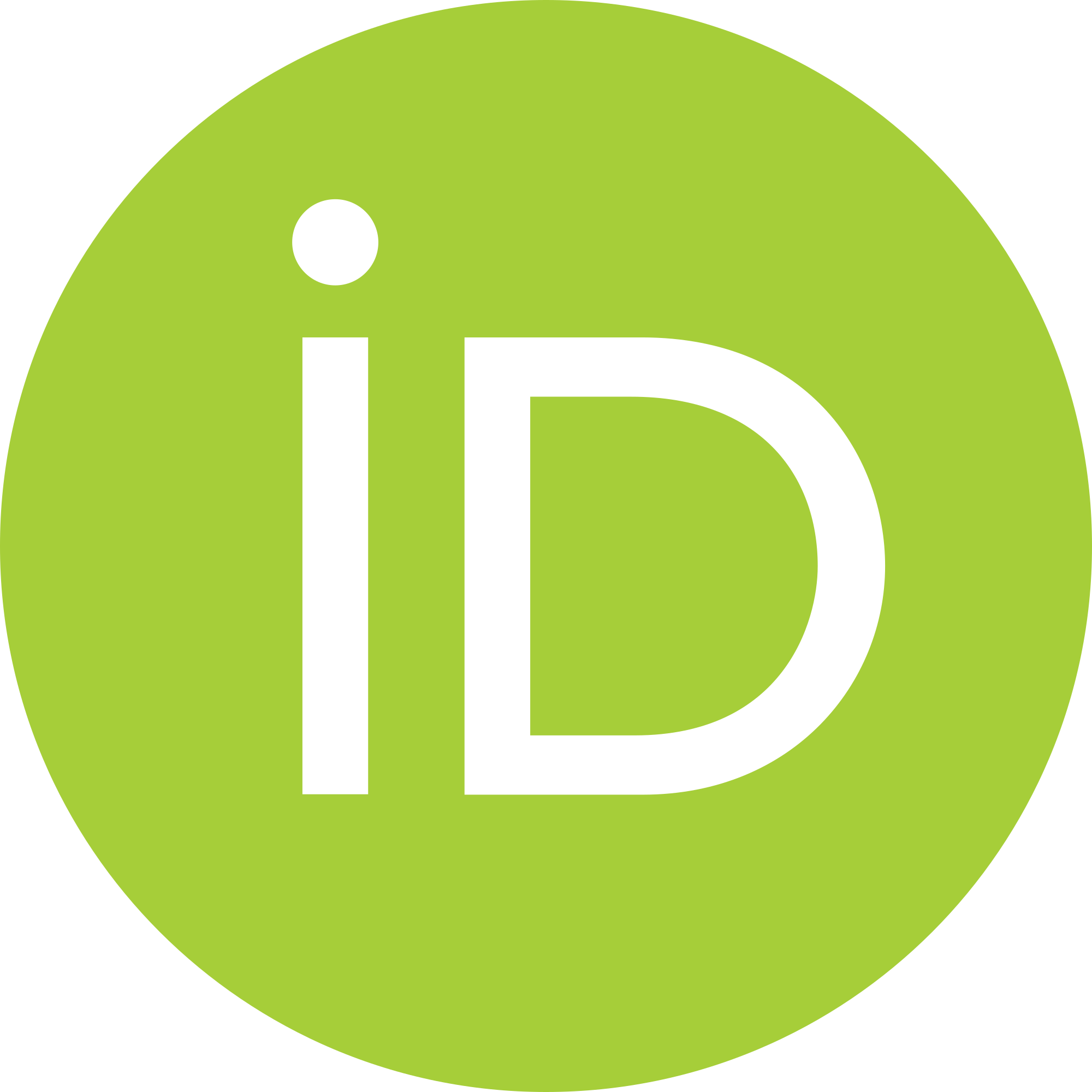}}}

\begin{document}

\title{\textsc{FastSpec}: Fast and versatile stellar spectrum synthesis code}

\author{Oleg Kochukhov\corrauth{oleg.kochukhov@physics.uu.se}\orcidlink{0000-0003-3061-4591}}

\authorrunning{Oleg Kochukhov} 

\institute{Department of Physics and Astronomy, Uppsala University, Box 524, SE-751~20 Uppsala, Sweden}

\date{Received 18 August 2026 / Accepted 14 October 2026}

\abstract
{Modelling data from modern spectroscopic surveys and time-resolved spectropolarimetric observations requires stellar synthetic spectra to be computed rapidly over wide wavelength ranges from line lists that increasingly reach hundreds of millions of transitions, and for stars spanning the entire Hertzsprung-Russell diagram. Spectropolarimetric analyses of magnetically active stars additionally demand a solution of the polarised radiative transfer problem.}
{We present \fastspec, a new local-thermodynamic-equilibrium (LTE) spectrum synthesis code designed around two goals: increased raw computational speed and broad astrophysical applicability, including the calculation of the full Stokes vector in the presence of a magnetic field.}
{\fastspec\ computes emergent intensity and flux spectra from plane-parallel and spherical model atmospheres. Chemical and ionisation equilibrium is obtained from the updated Spectroscopy Made Easy (\sme) equation-of-state package, continuum opacities follow the \atlas\ family of routines, hydrogen bound-bound and bound-free absorption uses the occupation-probability formalism, and line absorption is evaluated with fast table-based Voigt and Faraday-Voigt profile algorithms. The transfer equation is integrated with DELO-B\'ezier formal solvers in both scalar and Stokes-vector form. Speed is achieved through aggressive OpenMP parallelisation, adaptive per-line wavelength windowing, and a memory-capped block-processing scheme that lets arbitrarily large line lists be handled within a user-defined RAM budget.}
{The code synthesises spectra thousands of \AA\ wide from line lists of $10^{8}$ molecular and atomic transitions in under a minute on a modern multi-core laptop while reproducing benchmark calculations from established codes. We demonstrate the code's performance and versatility on the solar spectrum, on the near-infrared molecular bands of an M dwarf, on the four Stokes parameters of a magnetic star, and on the hydrogen lines of hot stars.}
{\fastspec\ combines the physical completeness of classical model atmosphere and spectrum synthesis codes with the speed required by large surveys and the polarised transfer needed for stellar magnetism studies into a single portable package.}


\maketitle
\nolinenumbers

\section{Introduction}
\label{sec:intro}

The comparison of observed stellar spectra with synthetic ones computed
from model atmospheres is the backbone of quantitative stellar
spectroscopy. It underlies the determination of atmospheric parameters
and chemical abundances of individual stars, the interpretation of data collected by large spectroscopic
surveys, and the mapping of stellar surface structures through
Doppler and Zeeman-Doppler imaging. A number of mature codes address
this spectrum synthesis problem in the local thermodynamic equilibrium (LTE) approximation, among them
\atlas/\synthe\ \citep{kurucz:2005}, \textsc{MOOG}
\citep{sneden:1973}, \turbospectrum\ \citep{plez:2012,gerber:2023},
\textsc{Spectrum} \citep{gray:1994}, \textsc{Synspec} \citep{hubeny:2017},
and Spectroscopy Made Easy \citep[\sme,][]{valenti:1996,piskunov:2017}. Codes that additionally
solve the polarised transfer equation for magnetic stars include
\textsc{Zeeman} \citep{landstreet:1988,wade:2001}, \textsc{MOOGStokes}
\citep{deen:2013}, \textsc{ZeeTurbo} \citep{cristofari:2023a}, \textsc{Synthmag} and
\textsc{Synmast} \citep{piskunov:1999,kochukhov:2007d,kochukhov:2010a}, and
the \textsc{Invers} family \citep{piskunov:2002a,kochukhov:2014} used for magnetic mapping.

Three developments have progressively raised the computational demands
placed on such codes. First, spectroscopic surveys now deliver observations for thousands
of stars so that even modest per-star costs aggregate into large
computing budgets, and iterative parameter determination multiplies the
number of syntheses required. Second, atomic and molecular line
databases have grown enormously. The \vald\ database
\citep{piskunov:1995,kupka:1999,ryabchikova:2015} contains several million
atomic transitions, while molecular line lists from the \exomol\
project \citep{tennyson:2024} routinely reach hundreds
of millions of lines per molecule, which must be processed to
model the spectra of cool stars, brown dwarfs, and exoplanets. Third,
spectropolarimeters attached to large telescopes produce time series of
all four Stokes parameters, whose modelling requires many polarised
syntheses per rotation phase. At the same time, robust interpretation of mean polarisation profiles
commonly employed in spectropolarimetry demands full Stokes parameter synthesis over broad
wavelength regions \citep{kochukhov:2026}. 
Recent codes such as \textsc{Korg}
\citep{wheeler:2023} have shown that careful algorithm design and modern
hardware can accelerate synthesis by large factors; the same
considerations motivate the present work.

In this paper we describe \fastspec, a spectrum synthesis code written
in modern Fortran and built explicitly for speed and versatility. Its
central design choices are (i) all computationally heavy loops -- over
wavelengths, depths, and lines -- are parallelised with OpenMP; (ii)
each spectral line contributes only within an adaptively determined
wavelength window so that the cost scales with the number of
significant line-wavelength interactions rather than with the
product of list size and grid size; (iii) the memory footprint is capped
by a user-defined limit, above which the line list is processed in
blocks, allowing lists of essentially unlimited size to be handled on
modest hardware; and (iv) profile functions are evaluated from
precomputed lookup tables. When a magnetic field and a line list with
Zeeman-splitting information are supplied, \fastspec\ solves the
polarised transfer problem and returns the full Stokes $IQUV$ spectrum.

The physical ingredients of \fastspec\ are, wherever possible, inherited
from well-tested public codes: the equation of state from \sme\
\citep{valenti:1996,piskunov:2017}, the continuum opacities from the
\atlas12 code \citep{kurucz:2005}, and the hydrogen line and
bound-free package \hbop/\textsc{Hlinop}
\citep{barklem:2003}. The originality of \fastspec\
lies in the way these ingredients are restructured, parallelised, and
combined with the adaptive windowing and memory management, which make
large-scale, magnetic synthesis fast.

The remainder of the paper is organised as follows.
Section~\ref{sec:overview} gives an overview of the code structure and
data flow. In Sect.~\ref{sec:physics} we describe the physical
foundations: the equation of state, continuum and hydrogen opacities,
line absorption and broadening, the Zeeman effect, and the radiative
transfer and disk-integration algorithms. Section~\ref{sec:performance}
presents the algorithmic ingredients responsible for the code's speed
together with performance benchmarks. In Sect.~\ref{sec:data} we summarise
the input data formats and in Sect.~\ref{sec:tests} validate the code
against reference calculations and illustrate its range of
applications. In Sect.~\ref{sec:summary} we summarise the paper. The input
keywords and output file formats are documented in the appendices.

\section{Code overview}
\label{sec:overview}

\begin{figure*}[t]
  \centering
  \includegraphics[width=0.92\textwidth]{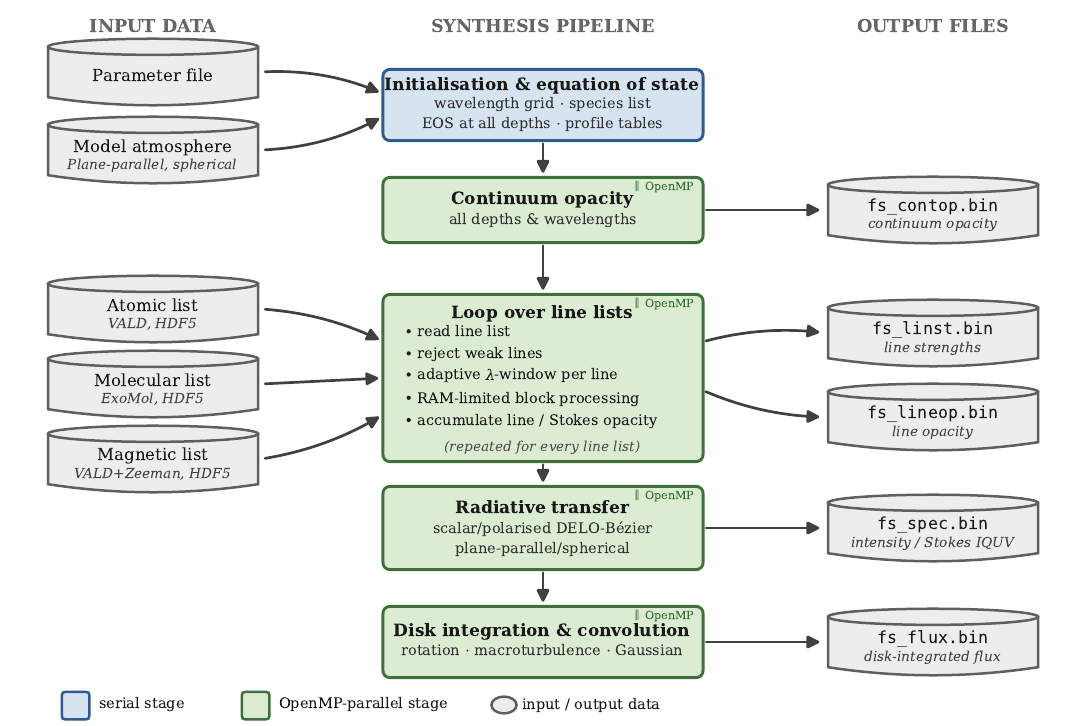}
  \caption{Schematic data flow of \fastspec. The input data (left) are the
    parameter file; the model atmosphere; and the atomic, molecular, and
    magnetic line lists. The output products (right) are the binary
    line-strength, line-opacity, continuum-opacity, intensity, and flux
    files. The central pipeline runs from top to bottom through the
    initialisation and equation of state, the continuum opacity, the loop
    over line lists (weak-line rejection, adaptive per-line wavelength
    windowing, and RAM-limited block processing), the scalar or polarised
    radiative transfer, and the disk integration and convolution. Green
    boxes mark the OpenMP-parallel stages, and the blue box indicates the (largely
    serial) initialisation.}
  \label{fig:flowchart}
\end{figure*}

\fastspec\ is written in Fortran~2008\footnote{See
\url{https://fortran-lang.org} for an overview of modern Fortran.} and
parallelised with OpenMP. This choice allows the well-tested routines of
established codes to be reused with minimal changes. Compiled Fortran code also remains considerably faster than
interpreted languages such as Python for computationally expensive tasks. The code is
organised into a small number of modules with clearly separated
responsibilities: shared data structures and physical constants; general
utilities; input and output; the equation of state; continuum, hydrogen, and
helium opacities; the line-absorption kernel and profile functions; the
radiative transfer solvers; and the disk-integration and convolution
routines. A high-level driver orchestrates these components in the
sequence shown in Fig.~\ref{fig:flowchart}.

A calculation proceeds through the following stages.
\begin{enumerate}
\item \emph{Initialisation.} The input parameter file is parsed, and the
  model atmosphere is read. The wavelength grid is constructed, either
  with a constant step in wavelength or a constant step in velocity
  (i.e.\ equidistant in $\ln\lambda$). The union of all chemical species
  appearing in the requested line lists, augmented by a fixed set of
  continuum absorbers, is assembled, and the equation of state is solved
  at every atmospheric depth. Lookup tables for the Voigt and, if needed,
  Faraday-Voigt functions are computed.
\item \emph{Continuum opacity.} The continuous absorption and scattering
  coefficients are computed for every depth and wavelength.
\item \emph{Line opacity.} The requested line lists are processed one at a
  time. For each list the lines are read, weak lines are rejected against
  a line-to-continuum threshold, an adaptive wavelength window is
  determined for every retained line, and the line opacity is accumulated
  onto the wavelength grid. Hydrogen lists and magnetic lists are handled
  by dedicated paths (Sects.~\ref{sec:hydrogen} and \ref{sec:zeeman}).
\item \emph{Radiative transfer.} The transfer equation is integrated for
  the requested set of limb angles $\mu$, yielding the emergent intensity
  spectrum (scalar case) or the emergent Stokes vector (magnetic case) at
  each $\mu$.
\item \emph{Disk integration and convolution.} If a flux spectrum is
  requested, the intensity spectra are disk-integrated with rotational
  broadening and radial-tangential macroturbulence, and optionally
  convolved with a Gaussian instrumental profile.
\item \emph{Output.} The selected products -- line-strength list, line
  and continuum opacities, intensity spectra, and flux spectrum -- are
  written as unformatted binary stream files.
\end{enumerate}

The code has a second mode of operation intended for interactive work.
Given a previously computed intensity-spectrum file and a set of
broadening parameters, it performs only the disk integration and
convolution described in step 5 above. This makes it inexpensive to explore the effect of
projected rotational velocity, macroturbulence, and spectral resolution
without recomputing the intensity spectra.

\section{Physical foundations}
\label{sec:physics}

\fastspec\ assumes a static atmosphere in LTE, in either
plane-parallel or spherical geometry.
The emergent spectrum is obtained by solving the transfer equation
along rays of direction cosine $\mu=\cos\theta$ through a model
atmosphere specified on a column-mass ($\varrho x$) depth scale. In this
section we describe, in the order in which they enter the calculation,
the equation of state, the opacity sources, the treatment of the Zeeman
effect, and the radiative transfer and disk-integration algorithms.

\subsection{Model atmospheres and the wavelength grid}
\label{sec:atmos}

Model atmospheres are read in the \texttt{.krz} 
format.\footnote{\url{https://marcs.oreme.org/documents/krzformat}} The header
specifies the effective temperature $\Teff$, the surface gravity $\logg$,
a model-type flag, the standard wavelength $\lambda_{\mathrm{std}}$ at
which reference opacities are tabulated, and a set of switches selecting
the continuum opacity sources. The header is followed by the chemical
composition, given as the abundances of the 99 elements from H to Es, and
by the depth-dependent structure: the column mass $\varrho x$, temperature
$T$, electron number density $n_{\mathrm e}$, total atomic number density
$n_{\mathrm a}$, and mass density $\rho$ at each of the $N_\tau$ depth
points. The format is directly compatible with the \atlas9\
\citep{kurucz:2005}, \textsc{Marcs} \citep{gustafsson:2008}, and \textsc{LLmodels}
\citep{shulyak:2004} models routinely used in the field.
A spherical model atmosphere is identified from its header, which then
also specifies the stellar radius and, as an additional depth column, the
geometrical height above the reference radius. These inputs define the extended
geometry in which the transfer equation is solved (Sect.~\ref{sec:rt}).

The default chemical composition can be overridden in two ways: a global
metallicity $[\mathrm{M/H}]$ can scale the solar abundance pattern of all
elements heavier than helium, or individual element abundances can be
specified explicitly. The reference solar abundances are those of
\citet{asplund:2021}, and the atomic weights follow the compilation
distributed with \turbospectrum.\footnote{\url{https://github.com/bertrandplez/Turbospectrum_NLTE}}

The wavelength grid is defined by its end points and a step that is
interpreted either as a constant wavelength increment (in \AA) or as a
constant velocity increment (in \kms), the latter giving a grid that is
uniform in $\ln\lambda$ and thus natural for radial-velocity and
rotational-broadening work. All calculations use air wavelengths; line lists 
supplied with vacuum wavelengths are converted to air when they are read.

\subsection{Equation of state}
\label{sec:eos}

Chemical and ionisation equilibrium is solved with the equation-of-state
package\footnote{\url{https://github.com/SpectroscopyMadeEasy/SMElib}}
of \sme\ \citep{valenti:1996,piskunov:2017}, incorporated with
small modifications and bug fixes. The package handles a default set of 497 species: the
neutral atoms and their first two ions for the 99 elements from H to Es
(only the first ion for H), about 190 molecules, and 15 negative ions,
plus free electrons. The equation of state itself supports up to six
ionisation stages (neutral to five times ionised), as required by the
high-ionisation continuum opacity package (Sect.~\ref{sec:contop}). 
Species that occur in the input line lists but not in the default
set are appended to the species list
automatically during the inventory pass over the lists
(Sect.~\ref{sec:data}) so that their number densities are available to the
line-opacity routines. The molecular
partition functions and equilibrium constants are based on the
compilation of \citet{barklem:2016}, supplemented by the data from \citet{sauval:1984} 
along with direct fits to \exomol\ partition functions for several molecules.

At each atmospheric depth, the code solves the coupled system of
particle-conservation equations for every element, the charge-neutrality
equation, and the total-pressure equation by Newton-Raphson iteration,
with an outer loop that enforces self-consistency of the electron
pressure. For temperatures above $12\,000$\,K, molecules are neglected and the
Saha equation is used directly; for cooler layers the full molecular
equilibrium is solved, using logarithmic partial pressures below
$1500$\,K for numerical stability. The solver returns, for every species,
the ratio $n_i/U_i$ of number density to partition function and the
partition function $U_i$ itself; the number density is recovered as their
product. This factorisation is convenient because the line-absorption
coefficient (Sect.~\ref{sec:lineop}) depends on $n_i/U_i$ directly. The
electron density, total particle density, and mass density returned by the
solver replace the values read from the model atmosphere, ensuring
thermodynamic consistency between the equation of state and the opacities.

\subsection{Continuum opacity}
\label{sec:contop}

The continuous absorption and scattering coefficients follow the
implementation in the \atlas12\ code of \citet{kurucz:2005}, translated to
Fortran~90 and reorganised so that all temperature-dependent quantities
that do not depend on wavelength are precomputed once per model, before
the parallel loop over wavelengths. The individual sources, each
controlled by a switch inherited from the model-atmosphere header, are
listed in Table~\ref{tab:contop}. They comprise bound-free and
free-free absorption by \ion{H}{i}, H$^-$, H$_2^+$, \ion{He}{i},
\ion{He}{ii}, and He$^-$; bound-free absorption by the astrophysically
important neutral and singly ionised metals (C, N, O, Mg, Al, Si, Ca,
Fe); photodissociation of CH and OH; collision-induced absorption by
H$_2$-H$_2$ and H$_2$-He pairs; the high-ionisation
metal opacities relevant for hot stars (up to five times ionised); Rayleigh scattering by
\ion{H}{i}, \ion{He}{i}, and H$_2$; and Thomson scattering by free
electrons. All number densities are taken from the equation of state, and
all bound-free and free-free terms include the stimulated-emission
factor $1-\mathrm{e}^{-h\nu/kT}$.

\begin{figure}[t]
  \centering
  \includegraphics[width=\hsize]{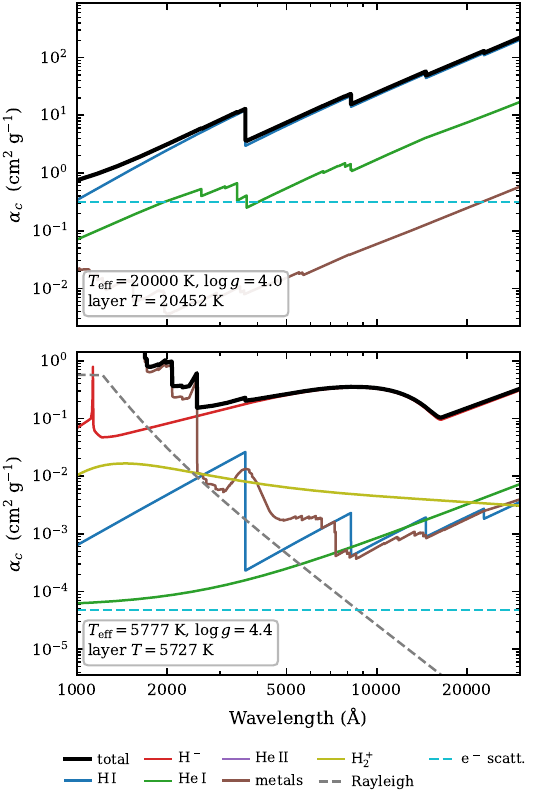}
  \caption{Continuum absorption coefficient per gram, $\alpha_c$, at a
    representative photospheric layer (temperature closest to $\Teff$)
    decomposed into the individual sources of Table~\ref{tab:contop} for a
    hot model (\emph{top}; $\Teff=20\,000$\,K, $\logg=4.0$) and the Sun
    (\emph{bottom}; $\Teff=5777$\,K, $\logg=4.44$). The black curve is the total absorption coefficient; the
    coloured curves are the isolated sources (solid: true absorption;
    dashed: Rayleigh and Thomson scattering), each computed by \fastspec\ with
    only the corresponding opacity switch active.
    }
  \label{fig:contop}
\end{figure}

Where the calculation also treats hydrogen lines with the
occupation-probability package (Sect.~\ref{sec:hydrogen}), the
bound-free contribution of the low-lying hydrogen levels is removed from
the continuum to avoid double counting.
The continuum is otherwise
identical whether or not a hydrogen line list is present.
Figure~\ref{fig:contop} illustrates the relative contributions of the
principal sources across the ultraviolet, optical, and near-infrared for a
hot ($\Teff=20\,000$\,K) and for a solar model, obtained directly from the
code by activating one source at a time.

\begin{table}[t]
\caption{Continuum opacity sources in \fastspec.}
\label{tab:contop}
\centering
\small
\begin{tabular}{l l}
\hline\hline
Process & Species / method \\
\hline
Bound-free, free-free & \ion{H}{i}, H$^-$, H$_2^+$ \\
Bound-free, free-free & \ion{He}{i}, \ion{He}{ii}, He$^-$ \\
Metal bound-free (cool) & \ion{C}{i}, \ion{Mg}{i}, \ion{Al}{i}, \ion{Si}{i}, \ion{Fe}{i} \\
Metal bound-free (warm) & \ion{C}{ii}, \ion{N}{i}, \ion{O}{i}, \ion{Mg}{ii}, \ion{Si}{ii}, \ion{Ca}{ii} \\
High-ion bound-free/free-free & C, N, O, Ne, Mg, Si, S, Fe \\
Photodissociation & CH, OH \\
Collision-induced abs. & H$_2$-H$_2$, H$_2$-He \\
Rayleigh scattering & \ion{H}{i}, \ion{He}{i}, H$_2$ \\
Thomson scattering & free electrons \\
\hline
\end{tabular}
\tablefoot{Each source can be switched on or off individually through the
opacity flags in the model-atmosphere header.}
\end{table}

\subsection{Hydrogen line and bound-free opacity}
\label{sec:hydrogen}

Hydrogen lines and the associated bound-free absorption are treated
jointly in the occupation-probability formalism, using the \hbop/%
\textsc{Hlinop} package\footnote{\url{https://github.com/barklem/hlinop}} 
of \citet{barklem:2003}, restructured for \fastspec\ so
that the level populations and all line-, depth-, and (line,\,depth)-%
dependent quantities are precomputed once for fast, thread-safe
evaluation over the wavelength grid. The formalism follows
\citet{dappen:1987} and \citet{hummer:1988}, with the
charged-perturber term treated as in \citet{hubeny:1994}. An occupation
probability $w_n$ is assigned to each bound level $n$. It describes the
gradual dissolution of the upper levels by the plasma microfield and
provides a physically consistent merging of the bound-bound and
bound-free opacity across the series limits so that the pseudo-continuum
between the highest resolved lines and the true continuum is captured
without discontinuities.

The Stark broadening of the hydrogen lines uses quasi-static Holtsmark
profiles following \citet{griem:1960} with corrections that approximate
the unified profiles of \citet{vidal:1973}, supplemented by impact electron broadening \citep{griem:1967}, 
a long-range normalisation correction \citep{stehle:1999}, and a Boltzmann factor on the red wing to approximate quantum effects \citep{stehle:1994}.
Self-broadening of the lower Balmer lines follows \citet{barklem:2000a}, resonance
broadening follows \citet{ali:1966}, and the Lyman\,$\alpha$ wing
includes the quasi-molecular H$_2$ and H$_2^+$ satellites of
\citet{allard:1994}. Two options are provided for the hydrogen bound-free
cross-section implementation: the tabulation used in \textsc{Marcs}
and the approximation from \atlas, with both relying on the same physical foundation
\citep{karzas:1961}. The two implementations agree to better than $1$\%.

The package returns, at every depth and wavelength, the total (line plus
continuum) and the pure continuum hydrogen opacity; their difference is
the bound-bound contribution, which is added to the line opacity, while
the continuum part is added to the continuum. Because the bound-free
part is computed here, it is switched off in the general continuum
routine (Sect.~\ref{sec:contop}) whenever a hydrogen line list is present.

\subsection{Line absorption}
\label{sec:lineop}

For a spectral line of a species with number density $n_i$ and partition
function $U_i$, lower-level energy $E_{\rm lo}$, and oscillator strength
$gf$, the monochromatic line-absorption coefficient per gram at the line
centre $\lambda_0$ is
\begin{equation}
\label{eq:lineop}
\alpha_0 = \frac{\sqrt{\pi}\,e^2}{m_{\mathrm e} c}\,
           \frac{gf}{U_i}\, n_i\,
           \mathrm{e}^{-E_{\rm lo}/kT}\,
           \bigl(1-\mathrm{e}^{-hc/\lambda_0 kT}\bigr)\,
           \frac{1}{\rho\,\Delta\nu_{\mathrm D}} ,
\end{equation}
where the Doppler width combines thermal and microturbulent motions,
\begin{equation}
\label{eq:doppler}
\frac{\Delta\nu_{\mathrm D}}{\nu_0}
   = \frac{\Delta\lambda_{\mathrm D}}{\lambda_0}
   = \frac{1}{c}\sqrt{\frac{2kT}{m_i}+\vmic^2}.
\end{equation}
The line-absorption coefficient at a wavelength offset $\Delta\lambda$ from the
line centre is $\alpha_\ell = \alpha_0\,H(a,\varv)$, where $H$ is the Voigt
function, $\varv =
\Delta\lambda/\Delta\lambda_{\mathrm D}$, and the damping parameter is
\begin{equation}
\label{eq:aparam}
a = \frac{\Gamma_{\rm tot}\,\lambda_0^2}{4\pi c\,\Delta\lambda_{\mathrm D}},
\qquad
\Gamma_{\rm tot} = \Gamma_{\rm rad}+\Gamma_{\rm Stark}+\Gamma_{\rm vdW}.
\end{equation}
The quantities in Eq.~(\ref{eq:lineop}) that depend only on the line, and
those that depend only on depth, are precomputed separately and combined
in the innermost loop so that adding a line at a new depth costs only a
few floating-point operations plus one Voigt evaluation.

The three damping constants are obtained as follows. The radiative width
$\Gamma_{\rm rad}$ is taken from the line list when available, otherwise
from the classical value \citep[e.g.][]{gray:2005}, $\Gamma_{\rm rad}=2.2233\times10^{15}/\lambda_0^2$,
with $\lambda_0$ in \AA. The quadratic Stark width $\Gamma_{\rm Stark}$
is taken from the line list or computed from the classical approximation
scaled with electron density, distinguishing neutral and ionic species
through the effective principal quantum number of the upper level
\citep{cowley:1971}; for the tabulated \ion{He}{i} lines it is replaced by
the values described in Sect.~\ref{sec:hestark}. The van der Waals width
$\Gamma_{\rm vdW}$ is taken from the line list, or from the Uns\"old
approximation \citep{unsold:1955} with an optional global scaling factor,
or -- when the line list provides the packed cross-section and
velocity-exponent parameters -- from the Anstee, Barklem, and O'Mara (ABO) theory of
\citet{anstee:1995} and \citet{barklem:2000b}. Perturbations by neutral
hydrogen and helium are included by default, and by molecular hydrogen
optionally. For molecular lines the Stark and van der Waals broadening is
switched off by default and can be enabled with a dedicated flag.

\subsubsection{Fast Voigt and Faraday-Voigt functions}
\label{sec:voigt}

Because a single synthesis may require $10^{10}$ or more profile
evaluations, the Voigt function $H(a,\varv)$ and its associated anomalous
dispersion (Faraday-Voigt) function $L(a,\varv)$, needed for polarised radiative
transfer, are evaluated from precomputed two-dimensional lookup tables
with bilinear interpolation. The tables are node-spaced uniformly in
$\varv$ and in $\sqrt a$, which concentrates resolution where the profile
varies fastest. Outside the tabulated domain, the functions are computed
from the asymptotic continued-fraction expansion of the Faddeeva
function. The tables themselves are built at start-up from a reference
implementation based on the trapezoidal expansion of
\citet{chiarella:1968}, accurate to $\sim$\,$10^{-12}$. A per-line caching scheme stores the
$a$-dependent interpolation state so that, when the Zeeman components of a
line are summed at a given depth, only the $\varv$-dependent part of the
lookup is repeated.

\begin{figure}[t]
  \centering
  \includegraphics[width=\hsize]{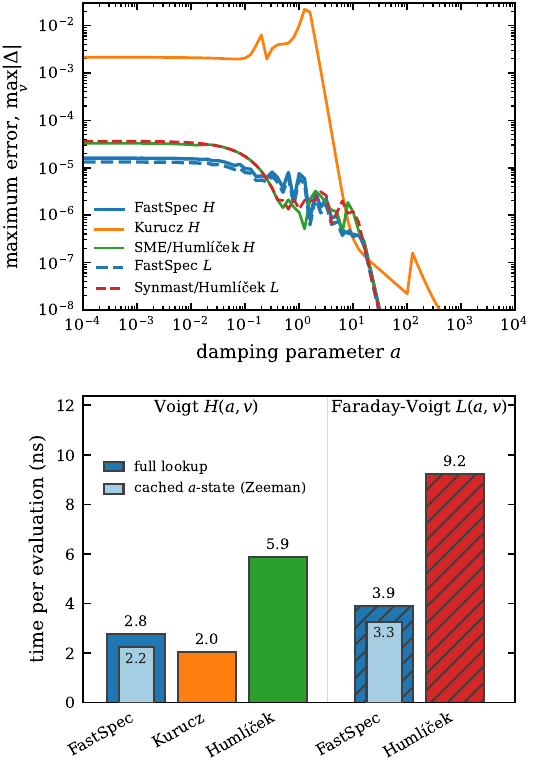}
  \caption{Accuracy and speed of the \fastspec\ table-based profile functions
    compared with other widely used algorithms. \emph{Top:} Maximum over
    $\varv$ of the absolute error of the Voigt profile $H(a,\varv)$ (solid
    lines) and the Faraday-Voigt profile $L(a,\varv)$ (dashed lines) as a
    function of the damping parameter $a$. 
    \emph{Bottom:} Corresponding time per evaluation
    on a single core. The light-blue bars
    over the \fastspec\ entries are the cached-$a$-state variants used when
    summing over the Zeeman components of a line at fixed depth.
    The bar heights are the mean time per call, printed above each bar; the
    left group refers to $H(a,\varv)$ and the right hatched group to
    $L(a,\varv)$. The \atlas/\synthe\ tabulation has no Faraday-Voigt
    counterpart and therefore appears only in the left group.
    }
  \label{fig:voigt}
\end{figure}

Figure~\ref{fig:voigt} compares the accuracy and speed of these table-based
functions with several algorithms in common use: the tabulated Voigt function
of \atlas/\synthe\ \citep{kurucz:2005}, also used by \textsc{Spectrum} \citep{gray:1994} 
and \textsc{Synspec} \citep{hubeny:2017}, the \sme\ implementation
\citep{piskunov:2017} of the \citet{humlicek:1982} $w_4$ rational approximation,
and the Huml\'{\i}\v{c}ek-based Faraday-Voigt routine of \textsc{Synmast}
\citep{kochukhov:2010a}. Over the full range of damping parameters, the
\fastspec\ tables reproduce the exact profile functions to better than
$2\times10^{-5}$ -- comparable to or better than the Huml\'{\i}\v{c}ek approximations,
and considerably more accurate than the \atlas/\synthe\ tabulation -- while
evaluating more than twice as fast as the former. 
Only the less accurate \atlas/\synthe\ tabulation is marginally faster. The cached-$a$-state
variant used in the Zeeman summation is faster still. We also examined the
analytic Voigt approximation used in the \textsc{Korg} code
\citep{wheeler:2023}; its maximum error closely follows that of the
\atlas/\synthe\ tabulation -- both peak at a few $\times10^{-2}$ near
$a\simeq1.3$ -- while its evaluation is slightly slower, so it is not shown
separately in Fig.~\ref{fig:voigt}.

\subsubsection{Helium Stark broadening}
\label{sec:hestark}

The strong \ion{He}{i} lines, which dominate the spectra of B-type stars,
are broadened predominantly by the linear and quadratic Stark effect and
are poorly described by the generic quadratic-Stark prescription above.
\fastspec\ therefore uses tabulated Stark widths and shifts for the
individual \ion{He}{i} lines, taken from the
calculations of \citet{dimitrijevic:1984,dimitrijevic:1990}. This
treatment closely follows that in \sme\ and related codes
\citep{valenti:1996,kochukhov:2010a} and is
reimplemented here for a larger set of \ion{He}{i} lines. A
line is matched to a table entry by its wavelength and lower-level energy,
and the depth-dependent width and line-centre shift are obtained by
interpolating the tabulated temperature dependence and scaling by the
electron, proton, and \ion{He}{ii} number densities from the equation of
state. The shift displaces the profile centre asymmetrically, which is
included both in the scalar and in the polarised opacity. These
per-depth widths and shifts are precomputed once during initialisation.
The same table-driven framework can be extended to any other species for
which detailed Stark calculations are available, for example \ion{Si}{i},
\ion{Cr}{i}, and \ion{Cr}{ii}
\citep{dimitrijevic:2003,dimitrijevic:2005,dimitrijevic:2007}.

\subsection{The Zeeman effect and polarised line opacity}
\label{sec:zeeman}

In the presence of a magnetic field of strength $B$, each line splits into
$\pi$ and $\sigma$ Zeeman components. \fastspec\ reads the full splitting
pattern -- the wavelength shifts and relative strengths of the $\pi$,
red-$\sigma$, and blue-$\sigma$ components -- from the magnetic line list
(Sect.~\ref{sec:data}) so that arbitrary Zeeman patterns, and partial
Paschen-Back patterns where provided, are handled exactly. 
The wavelength shift of a component scales as 
\begin{equation}
\Delta\lambda_B = 4.6686\times10^{-13} g\,\lambda_0^2 B,
\end{equation}
with $\lambda_0$ in \AA, $B$ in gauss, and the unscaled shift $g$ given in the input line list.

From the split components, the code builds the seven independent elements
of the $4\times4$ absorption-dispersion matrix. The three absorption
profiles $\eta_p,\eta_b,\eta_r$ (for the $\pi$, blue-$\sigma$, and
red-$\sigma$ groups) are summed over the Zeeman splitting patterns, with
each Voigt function scaled by the respective component strength and the line-centre opacity
coefficient $\alpha_0$ of Eq.~(\ref{eq:lineop}) and then accumulated over all
contributing lines. The three anomalous dispersion profiles $\rho_p,\rho_b,\rho_r$ are
the corresponding sums of Faraday-Voigt functions. In terms of the angle
$\gamma$ between the field and the line of sight and the field azimuth
$\chi$ in the plane of the sky, the elements of the propagation matrix are
given by the expressions \citep[e.g.][]{polarization:2004}
\begin{align}
\eta_I &= \tfrac12\!\left[\eta_p\sin^2\gamma
        + \tfrac12(\eta_b+\eta_r)(1+\cos^2\gamma)\right], \nonumber\\
\eta_Q &= \tfrac12\!\left[\eta_p-\tfrac12(\eta_b+\eta_r)\right]
              \sin^2\gamma \cos2\chi, \nonumber\\
\eta_U &= \tfrac12\!\left[\eta_p-\tfrac12(\eta_b+\eta_r)\right]
              \sin^2\gamma \sin2\chi, \\
\eta_V &= \tfrac12(\eta_r-\eta_b)\cos\gamma \nonumber
\end{align}
and analogously for the dispersion terms $\rho_{Q,U,V}$, with $\eta$
replaced by $\rho$. These profiles are accumulated on the wavelength grid
in the same adaptive, memory-limited manner as the scalar line opacity,
with the window width enlarged to cover the maximum Zeeman shift. The
non-magnetic lines and the continuum contribute only to the diagonal
absorption $\eta_I$.

\subsection{Radiative transfer}
\label{sec:rt}

In the scalar (non-magnetic) case, the emergent specific intensity at limb
angle $\mu$ and wavelength $\lambda$ is obtained by integrating
\begin{equation}
\mu\,\frac{\mathrm dI_\lambda}{\mathrm d\tau_\lambda}
   = I_\lambda - S_\lambda
\end{equation}
along the ray, where the monochromatic optical depth $\tau_\lambda$ is
built from the total (line plus continuum) absorption on the column-mass
scale and the source function $S_\lambda$ is the Planck function. The
integration uses the DELO-B\'ezier formal solver developed by
\citet{de-la-cruz-rodriguez:2013}. In their implementation the optical-depth increments are estimated
with a monotonic B\'ezier quadrature, and the source function is
interpolated with B\'ezier polynomials whose control points are set by
monotonicity-preserving derivatives, which suppresses the spurious
oscillations of higher-order schemes. Short-characteristic coefficients
are evaluated from stable analytic expansions at small optical-depth steps.

In the magnetic case, the scalar equation is replaced by the vector
transfer equation for the Stokes vector
$\mathbf I = (I,Q,U,V)^{\mathsf T}$,
\begin{equation}
\mu\,\frac{\mathrm d\mathbf I}{\mathrm d\tau}
   = \mathbf K\,(\mathbf I - \mathbf S),
\label{eq:prt}
\end{equation}
where $\mathbf S = (S,0,0,0)^{\mathsf T}$ and $\mathbf K$ is the
propagation matrix built from the $\eta$ and $\rho$ elements described in
Sect.~\ref{sec:zeeman},
\begin{equation}
\label{eq:kmatrix}
\mathbf K = \frac{1}{\eta_I}
\begin{pmatrix}
 \eta_I  &  \eta_Q  &  \eta_U  &  \eta_V \\
 \eta_Q  &  \eta_I  &  \rho_V  & -\rho_U \\
 \eta_U  & -\rho_V  &  \eta_I  &  \rho_Q \\
 \eta_V  &  \rho_U  & -\rho_Q  &  \eta_I
\end{pmatrix} .
\end{equation}
Here $\eta_I$ is the total of the magnetic-line
diagonal absorption, the non-magnetic line opacity, and the continuum, and
$\eta_{Q,U,V}$ and $\rho_{Q,U,V}$ are the polarised-absorption and
anomalous dispersion coefficients. The common factor $1/\eta_I$ multiplies
every element of the propagation matrix so that its diagonal reduces to unity.
Equation~(\ref{eq:prt}) is then integrated
with the vector version of the DELO-B\'ezier solver \citep{de-la-cruz-rodriguez:2013}, which
requires the inversion of a $4\times4$ matrix per depth step, 
performed by Gaussian elimination with partial pivoting. 
The two solvers share the same optical-depth and derivative
routines. Both are validated in Sect.~\ref{sec:tests}.

For the magnetic geometry, the field can be specified either directly in
the observer's frame, as modulus, inclination to the line of sight, and
azimuth -- appropriate for a uniform field seen at a fixed aspect -- or
in the stellar surface frame as the radial, meridional, and azimuthal
components, in which case \fastspec\ projects the field onto the line of
sight separately at each limb angle before accumulating line opacities.
The observer-frame specification also accepts an individual inclination
and azimuth for each limb angle so that an arbitrary 
set of field orientations computed externally 
can be supplied directly.

The solvers described above assume a plane-parallel geometry. For
spherically extended atmospheres, \fastspec\ additionally implements the
spherical formal solution of \sme\ \citep{piskunov:2017}, applying the same
DELO-B\'ezier short-characteristic integration along the parallel rays that
sample the curved atmosphere, in both the scalar and the polarised case.

\subsection{Disk integration and convolution}
\label{sec:convol}

Flux spectra are obtained by integrating the emergent intensity over the
stellar disk. The disk is divided into concentric annuli centred on the
$\mu$ points; by default these are spaced so as to sample equal projected
areas. Each annulus is weighted by its fractional projected area. Within
each annulus, rigid rotation is treated by convolving the local intensity
with the analytic distribution of line-of-sight velocities appropriate to
an annulus of finite width, which avoids the artefacts of a
pixel-sampled rotation kernel \citep{valenti:1995}. Radial-tangential macroturbulence
\citep{gray:2005} is applied as a convolution with a $\mu$-dependent kernel
that combines the radial and tangential Gaussian components in equal
proportion. A final convolution with a Gaussian approximates the effect of an
instrumental profile; its width may be given directly or as a resolving
power. In the magnetic case, the same broadening is applied to all four
Stokes parameters. The disk integration, like the opacity and
radiative-transfer stages, is OpenMP-parallel.

\section{Performance and algorithmic design}
\label{sec:performance}

The speed of \fastspec\ is derived from four main ingredients that we 
describe in this section. This description is followed by benchmark measurements.

\subsection{Adaptive line windowing}
\label{sec:windowing}

\begin{figure}[!th]
  \centering
  \includegraphics[width=\hsize]{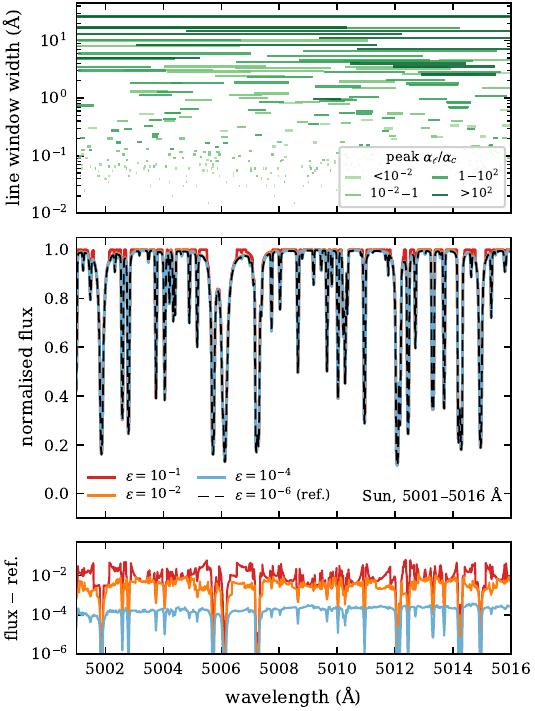}
  \caption{Adaptive line windowing and the line-to-continuum opacity threshold
    $\varepsilon$ illustrated on a segment of a solar synthetic spectrum (\vald\ `extract all' atomic list,
    \textsc{Marcs} solar model, no macroscopic broadening applied).
    \emph{Top:} Individual accumulation window of every line retained in the
    5001--5016\,\AA\ segment, each drawn as a horizontal bar spanning
    $[\lambda_{\rm left},\lambda_{\rm right}]$, positioned vertically at its
    width and coloured by the line's peak line-to-continuum opacity ratio
    ${\rm max}(\alpha_\ell/\alpha_c)$.
    \emph{Middle:} Corresponding normalised flux spectrum
    computed with $\varepsilon=10^{-1}$, $10^{-2}$ and the default $10^{-4}$,
    with a converged reference ($\varepsilon=10^{-6}$, dashed), which
    nearly coincides with the $\varepsilon=10^{-4}$ curve.
    \emph{Bottom:} Difference from the reference, with the same
    $\varepsilon$ colours as in the middle panel.
    }
  \label{fig:window}
\end{figure}

\begin{figure}[!th]
  \centering
  \includegraphics[width=\hsize]{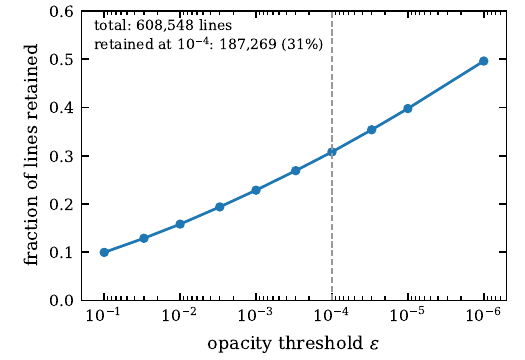}
  \caption{Fraction of the full (`extract all') \vald\ atomic line list
    retained by the weak-line rejection as a function of the opacity threshold
    $\varepsilon$ for the solar synthesis of Fig.~\ref{fig:window}. The dashed
    line marks the default $\varepsilon=10^{-4}$.}
  \label{fig:retention}
\end{figure}

The cost of a naive synthesis scales as the product of the number of
lines and the number of wavelength points, because every line is in
principle evaluated at every wavelength. In reality a line contributes
measurably only over a narrow interval around its centre. \fastspec\
determines this interval for every line before accumulating its opacity.
The central line-to-continuum opacity ratio is first computed; if it lies
below a user-controlled threshold $\varepsilon$ (typically $10^{-4}$), the
line is discarded altogether. For a retained line, the window half-width
is obtained in closed form by inverting the Voigt profile at the
threshold, evaluated at each atmospheric depth and maximised over depth,
giving a pair of grid indices that bracket the line without any profile
evaluation or iterative search. In the hydrogen and magnetic paths, where
the profile is not a single Voigt function, the window is instead found
numerically, by growing a trial offset geometrically until the ratio
drops below $\varepsilon$ and refining the boundaries by bisection. Because the window is set
individually for each line, its width varies enormously -- from a few tens of
m\AA\ for the weakest retained lines to tens of \AA\ for the strongest lines -- as
shown in the top panel of Fig.~\ref{fig:window}. Each line is shown as a
single bar, spanning its accumulation interval and placed at its width, so the
vertical spread at any wavelength shows how differently the windows of
neighbouring lines are sized. The remaining panels
illustrate the effect of the threshold on a solar synthesis with a full (`extract all') \vald\
atomic line list. At the default $\varepsilon=10^{-4}$,
only about a third of the lines are retained (Fig.~\ref{fig:retention}), yet the emergent flux spectrum is
indistinguishable from one computed with a threshold two orders of magnitude
smaller, whereas a coarse $\varepsilon=10^{-1}$ visibly suppresses the weaker
blends. Weak-line rejection thus removes the majority of a complete line list at
essentially no cost in accuracy.

\subsection{Memory-limited block processing}
\label{sec:ram}

\begin{figure}[!t]
  \centering
  \includegraphics[width=\hsize]{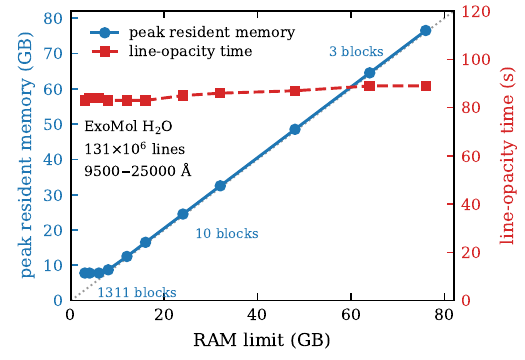}
  \caption{Memory-limited block processing illustrated on the \exomol\
    H$_2$O line list ($131\times10^6$ lines in the $9500$--$25\,000$\,\AA\
    range, \textsc{Marcs} $T_\mathrm{eff}=3500$\,K, $\logg=5.0$ model) on a
    128\,GB machine. The figure shows peak resident memory (solid, left axis) and
    line-opacity computation time (dashed, right axis) as a function of the
    user-defined RAM cap.
    }
  \label{fig:memory}
\end{figure}

Storing the per-line, per-depth central opacities and profile parameters
for a very large list would exceed available memory. \fastspec\ therefore
caps the memory devoted to these arrays at a user-controlled limit and, when
a list is larger than this cap allows, processes it in contiguous blocks
of lines. The wavelength range is partitioned into segments such that all
lines contributing to a segment fit within one block; the code then loads
one block of per-line data at a time, accumulates its contribution over
the corresponding wavelength segment, and moves on. Lines whose opacity
windows are exceptionally wide are excluded from this segmentation and
accumulated individually over their windows, preventing a single broad line
from inflating a segment beyond the block size and from slowing the
per-wavelength accumulation. In this way lists of
essentially unlimited length are handled within a fixed memory footprint,
at the cost of recomputing the per-line data for lines that straddle
segment boundaries. The maximum block size is derived from the requested
RAM limit and the dimensions of the problem. Figure~\ref{fig:memory}
shows the result of computations with the \exomol\ water list on a 128\,GB machine. The peak
memory closely follows the requested cap over more than a decade in memory,
down to an irreducible floor (${\sim}8$\,GB for this large H$_2$O line list) set by the permanently
resident line list and the spectral and index arrays. Over this whole
range, the run time is essentially constant: splitting the list into as few
as three or as many as ${\sim}1300$ blocks changes the line-opacity time by
only a few percent, because the lines recomputed at segment boundaries are
a negligible fraction of the total. Block processing therefore renders the
memory footprint an essentially free parameter; indeed for this list the
single-block working set (${\sim}170$\,GB) exceeds even a 128\,GB machine,
so the scheme is not merely convenient but necessary.

\subsection{Parallelisation and profile lookups}
\label{sec:openmp}

All heavy loops -- the continuum opacity over wavelength, the per-line
setup and windowing over lines, the opacity accumulation over wavelength,
the radiative transfer over wavelength, and the convolutions -- are parallelised
with OpenMP. The precomputation of the depth-only and line-only parts of
the opacity kernel, together with the shared read-only Voigt tables,
ensures that the parallel regions carry no synchronisation overhead beyond
the loop scheduling. Combined with the fast table-based profiles
(Sect.~\ref{sec:voigt}), this yields near-linear scaling over the first
$\sim$\,10 cores. At higher thread counts, the opacity accumulation becomes
limited by memory bandwidth rather than by synchronisation, as quantified in
Sect.~\ref{sec:benchmarks}.

\begin{figure}[!t]
  \centering
  \includegraphics[width=\hsize]{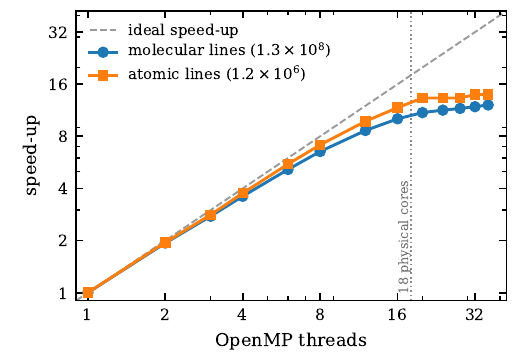}
  \caption{Strong-scaling speed-up $S(N)=T(1)/T(N)$ of the line-opacity
    computation versus the number of OpenMP threads for two
    very different line lists: the \exomol\ molecular water list
    ($1.3\times10^{8}$ lines, cool-dwarf \textsc{Marcs} model) and the solar
    atomic list ($1.2\times10^{6}$ lines). The dashed line is the ideal linear
    speed-up, and the dotted vertical line marks the 18 physical cores of the
    Intel Xeon\,W iMac Pro workstation used for this test (threads beyond it are logical).}
  \label{fig:scaling}
\end{figure}

Figure~\ref{fig:scaling} characterises the parallel scaling of \fastspec\ for
the line-opacity computation, which dominates a large synthesis, on a
workstation with 18 physical cores (36 logical threads with simultaneous
multithreading). For the \exomol\ water list (131 million lines,
computed for a cool-dwarf \textsc{Marcs} model)
over the 9500--25\,000\,\AA\ range, the scaling is near-linear
through the first ten threads (90\% parallel efficiency on four threads,
82\% on eight) and then rolls off, reaching a speed-up of $12\times$ on 36
threads. A very different workload -- the solar atomic list, with full
pressure broadening and the hydrogen lines treated by the occupation-probability
package but some two orders of magnitude fewer lines -- follows a nearly
identical curve, scaling marginally better and reaching $14\times$. The
near-coincidence of the two curves places the origin of the roll-off in the
hardware rather than in the algorithm or the line list: the opacity
accumulation has a low arithmetic intensity, streaming the precomputed
per-line, per-depth profile coefficients from memory for only a cached
Voigt-table lookup and a multiply-add per wavelength point, so once more than
about ten cores are active, they saturate the shared memory bandwidth, and
further threads increasingly stall waiting for data rather than computing. The
final doubling from 18 to 36 threads adds only logical (hyperthreaded) cores,
which share physical execution units and memory ports and barely improve a
bandwidth-bound kernel. The atomic list carries more arithmetic per line and is
correspondingly less bandwidth-bound, which is why it attains the slightly
higher speed-up.

\subsection{Benchmarks}
\label{sec:benchmarks}

Table~\ref{tab:timing} lists representative wall-clock times for several
characteristic spectrum synthesis problems, measured on a 14-core MacBook M4 Pro
laptop using all cores. For each run we separate the total wall-clock time
$T_{\rm all}$ into the atomic and molecular line-opacity accumulation
$T_{\rm opac}$ (the dominant cost for large lists, excluding the
hydrogen-line package) and the radiative transfer $T_{\rm RT}$; the
remainder covers the equation of state, continuum opacity, hydrogen lines,
and I/O. 
In all cases, spectra were calculated at seven limb angles using a 0.3\,\kms\ velocity step,
corresponding to a wavelength sampling of $\lambda/\Delta\lambda\approx10^6$.
Atomic syntheses over the $3000$--$25\,000$\,\AA\ range using the full (`extract all') \vald\ data,
whether for the Sun or for hot stars, complete in about ten seconds.
Their corresponding radiative transfer -- unpolarised, seven $\mu$ angles, over two million
wavelength points -- takes less than two seconds. Adding the full set of
solar molecular lists raises the opacity time to $7$\,s and the total to
$13$\,s but leaves the transfer unchanged. The two largest cases are dominated by the \exomol\ water list:
even with ${\sim}10^{8}$ contributing lines, the cool-dwarf synthesis
completes in about forty seconds for $\approx$\,$10^6$ wavelengths, and the full polarised $IQUV$ synthesis
in under fifty seconds for the same wavelength grid. The polarised transfer step itself stays cheap
($3.0$\,s), scaling with the number of wavelength points rather
than the line count, so most of the additional cost is the Zeeman treatment
of the magnetic-line opacity. 

These tests demonstrate that combined atomic and molecular line lists of hundreds of
millions of lines can be synthesised over the full optical-infrared range
in under a minute on a single multi-core machine. A direct
comparison of the speed and memory usage of \fastspec\ with the modern,
multi-threaded \textsc{Korg} code is presented in Appendix~\ref{app:korg}.

\begin{table*}[t]
\caption{Representative wall-clock times.}
\label{tab:timing}
\centering
\small
\begin{tabular}{l c c c c c c c}
\hline\hline
Model / line list & Range (\AA) & $N_\lambda$ & $N_{\rm lines}$ & $N_{\rm mag}$ & $T_{\rm all}$ (s) & $T_{\rm opac}$ (s) & $T_{\rm RT}$ (s) \\
\hline
Sun / atomic          & 3000--25\,000 & $2.12\times10^{6}$ & $3.53\times10^{5}$  & \dots & \phantom{0}6.9 & \phantom{0}1.0 & \phantom{0}1.3 \\
Sun / atomic + molecular         & 3000--25\,000 & $2.12\times10^{6}$ & $1.12\times10^{6}$ & \dots  & 13.0 & \phantom{0}6.9 & \phantom{0}1.3 \\
t20000g4.0 / atomic          & 3000--25\,000 & $2.12\times10^{6}$ & $3.86\times10^{4}$ & \dots  & 10.7 & \phantom{0}0.5 & \phantom{0}1.5 \\
t9000g4.0 / atomic          & 3000--25\,000 & $2.12\times10^{6}$ & $2.04\times10^{5}$  & \dots & 11.0 & \phantom{0}0.8 & \phantom{0}1.6 \\
t9000g4.0 / Stokes $IQUV$ & 3000--25\,000 & $2.12\times10^{6}$ & $2.04\times10^{5}$  & $2.04\times10^{5}$  & 24.7 & \phantom{0}8.0 & \phantom{0}8.2 \\
t3500g5.0 / atomic + molecular & 9500--25\,000 & $9.67\times10^{5}$ & $9.67\times10^{7}$ & \dots  & 43.0 & 40.1 & \phantom{0}0.6 \\
t3500g5.0 / Stokes $IQUV$ & 9500--25\,000 & $9.67\times10^{5}$ & $9.67\times10^{7}$ & $4.74\times10^{4}$  & 48.5 & 43.2 & \phantom{0}3.0 \\
\hline
\end{tabular}
\tablefoot{These test calculations were executed on a MacBook M4 Pro (14 cores: ten performance
and four efficiency, 48\,GB RAM) using all cores. $N_\lambda$ is the number of wavelength points,
$N_{\rm lines}$ is the number of lines
retained above the opacity threshold, and $N_{\rm mag}$ is the number treated
with the full polarised transfer. $T_{\rm all}$ is the total
wall-clock time of the run, $T_{\rm opac}$ is the atomic and molecular
line-opacity accumulation (the total line-opacity time excluding the
hydrogen-line package), and $T_{\rm RT}$ is the (polarised, for the Stokes
$IQUV$ cases) radiative transfer.
}
\end{table*}

\section{Input data and line lists}
\label{sec:data}

\fastspec\ reads three kinds of line list. A single synthesis can
combine atomic, molecular, and magnetic line data.

Atomic lists are supplied as HDF5 files containing, for each
transition, the wavelength, element and ionisation stage, lower-level
energy, oscillator strength, and the radiative, Stark, and van der Waals
damping constants. These files can be produced by converting \vald\
extractions. For the calculations in this paper, we used a single
`extract all' file packing the entire \vald\ atomic data in a 36\,MB HDF5 file. 
A list containing only hydrogen lines
is recognised automatically and routed to the occupation-probability
package (Sect.~\ref{sec:hydrogen}); it must be supplied first. The reader
uses a two-pass strategy, first taking an inventory of the species present
to build the equation-of-state species list, then reading the transitions
that fall within the requested wavelength range.

Molecular lists are single-species HDF5 files, such as conversions
of \exomol\ \citep{tennyson:2024}, containing the
wavelength, lower-level energy, oscillator strength, upper-level
statistical weight, and radiative width. The isotopologue is identified
from a numerical species code, and the oscillator strengths are scaled by
the appropriate isotopic fraction so that a molecule can be synthesised
from a set of per-isotopologue lists. 

Magnetic lists are ASCII files in the format used by the
\textsc{Synthmag} and \textsc{Synmast} codes
\citep{piskunov:1999,kochukhov:2010a}. In addition to the usual
atomic or molecular line data in \vald\ `extract stellar' format, they contain, for each line, the explicit
Zeeman splitting pattern: the number, shifts, and strengths of the $\pi$
and $\sigma$ components. This allows anomalous Zeeman and, where tabulated,
partial Paschen-Back patterns to be modelled exactly.
As an alternative to this precomputed form, the magnetic line data can be
supplied in an HDF5 line list that stores, in addition to the usual atomic
quantities, the lower- and upper-level angular momenta $J$ and Land\'e
factors $g$. \fastspec\ then computes the anomalous Zeeman splitting pattern
of each line internally as the list is read. Lines lacking Land\'e
factor information are treated as non-magnetic. A magnetic list triggers
the polarised transfer only when a magnetic field is also specified;
otherwise it is synthesised as an ordinary line list.

\fastspec\ is driven by an input file that specifies the wavelength range and grid type, 
the line lists and model atmosphere files to use, the desired outputs, disk integration and convolution parameters, 
elemental abundances, and optional computational settings.
The full set of input keywords is documented in
Appendix~\ref{app:input}, and the binary output formats in
Appendix~\ref{app:output}.

\begin{figure*}[!th]
  \centering
  \includegraphics[width=\textwidth]{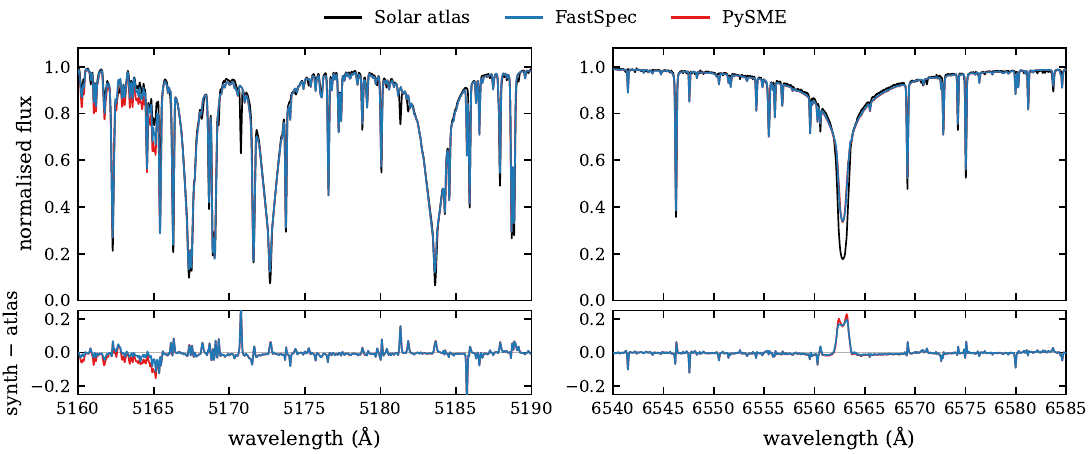}
  \caption{Solar flux spectrum in the \ion{Mg}{i}\,$b$ triplet region
    (\emph{left}) and around H$\alpha$ (\emph{right}). \emph{Upper panels:} \fastspec\ (blue)
    and an independent \pysme\ synthesis (red) computed from the same
    1D \textsc{Marcs} solar model, solar abundances \citep{asplund:2021}, and Gaia-ESO line list \citep{heiter:2021}
    are shown together with the
    solar flux atlas \citep{wallace:2011} (black). \emph{Lower panels:}
    Corresponding synthetic-minus-observed residuals.}
  \label{fig:sun}
\end{figure*}

\section{Tests and applications}
\label{sec:tests}

We illustrate the correctness and versatility of \fastspec\ with four
applications. Together, these tests span most of the parameter space for which the
code is intended. 

\paragraph{The solar spectrum.} Figure~\ref{fig:sun} compares synthetic
solar flux spectra with the observed, telluric-corrected solar flux atlas
of \citet{wallace:2011} in two diagnostic windows: the \ion{Mg}{i}\,$b$
triplet and H$\alpha$. The synthesis uses a \textsc{Marcs} solar model
($\Teff=5777$\,K, $\log g=4.44$), 
a microturbulence of $0.8$\,\kms, the built-in solar abundances of
\citet{asplund:2021}, and the Gaia-ESO survey line list \citep{heiter:2021}
-- its atomic lines together with the C$_2$ and MgH bands that blend into
the $b$-triplet region. The emergent flux is disk-integrated over seven
limb angles and broadened by a
radial-tangential macroturbulence of $3.2$\,\kms\ and a projected rotational velocity of
$1.9$\,\kms. To provide an independent check, the same model atmosphere,
abundances, line list, and broadening are also passed to \pysme\
\citep{piskunov:2017,jian:2026}.\footnote{\url{https://github.com/SpectroscopyMadeEasy/PySME}}

\begin{figure*}[t]
  \centering
  \includegraphics[width=\textwidth]{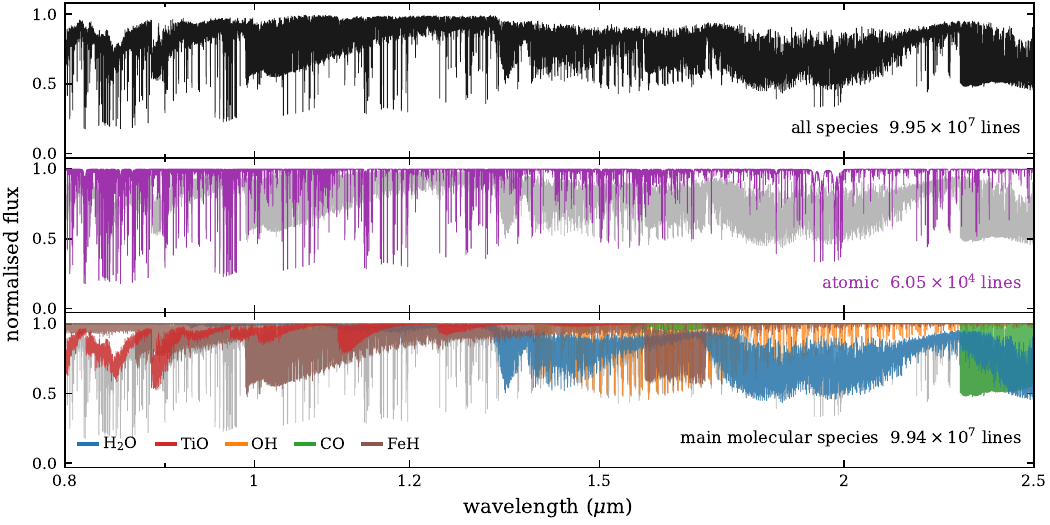}
  \caption{Synthetic near-infrared spectrum of a cool M-dwarf model
    ($\Teff=3500$\,K, $\log g=5.0$). 
    \emph{Top:} Complete
    spectrum including every opacity source of the input list.
    \emph{Middle:} Atomic-line contribution alone (magenta) overplotted on the full spectrum (grey).
    \emph{Bottom:} Main individual molecular contributions -- H$_2$O, TiO,
    OH, CO, and FeH -- each computed with only that species active and
    overplotted in colour. 
    }
  \label{fig:mdwarf}
\end{figure*}

The two codes agree closely across both windows, matching each other line
by line and in the cores and wings of the strong features, including H$\alpha$.
Both reproduce the observed metal lines and the \ion{Mg}{i}\,$b$
triplet well. Two differences stand out. The synthetic H$\alpha$ core is
shallower than observed because it forms in the uppermost photosphere and
chromosphere, which a static one-dimensional LTE model does not describe;
the mismatch therefore reflects the model atmosphere rather than the
spectral synthesis. Near the C$_2$ Swan bandhead at $5165$\,\AA, the two
syntheses diverge slightly, \pysme\ producing marginally stronger C$_2$
absorption. The difference amounts to about $2$\% of the continuum flux
across the band, reaching $7$\% at the head near $5165$\,\AA, while redward of
it the two codes agree to better than $0.6$\%. As both read identical line data, this traces back to
a slightly different C$_2$ dissociation energy adopted for the
molecular equilibrium calculations in the two codes. The test simultaneously
exercises the continuum opacity, the equation of state, and the atomic and
molecular line opacity and broadening, and cross-validates them against an
independent implementation.

\paragraph{Molecular bands in an M dwarf.} Figure~\ref{fig:mdwarf} shows
the complete near-infrared spectrum ($0.8$--$2.5\,\mu$m) of a cool
($\Teff=3500$\,K, $\log g=5.0$) dwarf, synthesised from the full input list
of atomic and molecular lines: the atomic list together with the CH, CN,
CO, CrH, FeH, HF, NaH, OH, TiO, and H$_2$O molecular lists. About
$10^{8}$ lines are retained above the opacity threshold,
overwhelmingly dominated by the \exomol\ H$_2$O list. To
identify the individual contributors, the spectrum is also recomputed with
only the atomic lines (middle panel) and with each of the main molecular
species in turn. The far red is
shaped by the TiO bands (here from five titanium isotopologues, $^{46}$TiO
to $^{50}$TiO), while H$_2$O dominates the near-infrared, together with the
CO first-overtone bandhead at $2.3\,\mu$m, the OH bands longward of
$1.5\,\mu$m, and FeH contributions around 1.0 and $1.6\,\mu$m. 
This test demonstrates the handling of very large molecular
line lists and the isotopic scaling of oscillator strengths, and stresses
the memory-limited block processing of Sect.~\ref{sec:ram}.

\begin{figure*}[!th]
  \centering
  \includegraphics[width=\textwidth]{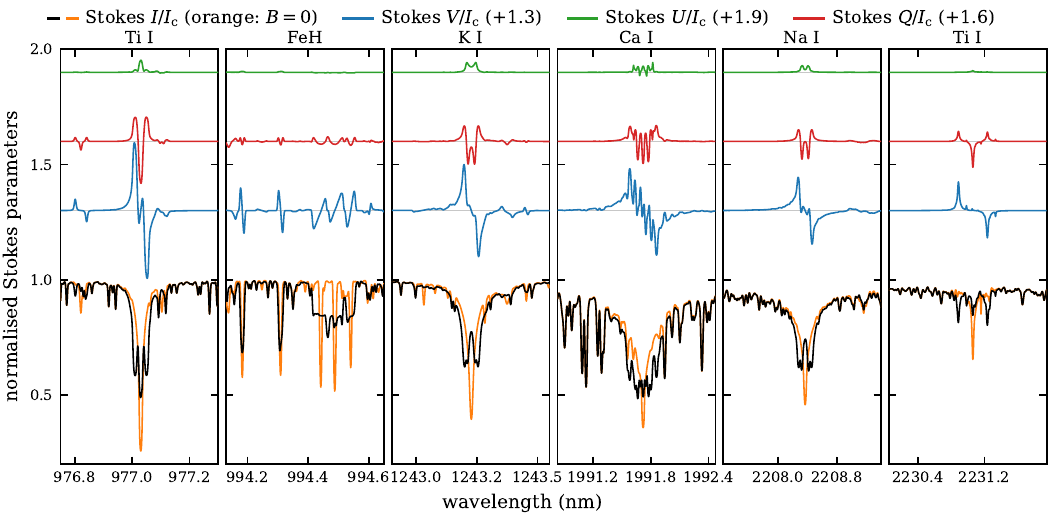}
  \caption{Polarised disk-centre near-infrared spectrum of a magnetic M-dwarf model
    ($\Teff=3500$\,K, $\log g=5.0$) with a uniform $3$\,kG field inclined by $45\degr$
    to the line of sight (zero azimuth). The six windows are each centred on a selected feature and labelled by the dominant
    species. In every panel the magnetic Stokes $I/I_{\rm c}$ (black) is drawn over the non-magnetic ($B=0$)
    profile (orange); the polarised profiles $Q/I_{\rm c}$, $U/I_{\rm c}$, and $V/I_{\rm c}$ have been
    shifted upward by the constant given in the legend.}
  \label{fig:stokes}
\end{figure*}

\paragraph{Polarised spectra of a magnetic M dwarf.}
Figure~\ref{fig:stokes} illustrates the polarised mode of \fastspec\ on the same
cool dwarf as Fig.~\ref{fig:mdwarf} ($\Teff=3500$\,K, $\log g=5.0$), now with
a uniform $3$\,kG field inclined $45\degr$ to the line of sight with zero
azimuth. The whole $0.95$--$2.5\,\mu$m interval is synthesised in a single run with
atomic and FeH lines \citep{shulyak:2010c} treated as polarised and other molecular species as non-magnetic.
Six narrow windows are extracted to
show the disk-centre Stokes $I$, $Q$, $U$, and $V$ profiles of representative
features: the atomic lines of \ion{Ti}{i}, \ion{K}{i}, \ion{Ca}{i}, and
\ion{Na}{i}, and the FeH lines within the FeH Wing-Ford band near $0.99\,\mu$m.

The comparison of the magnetic and field-free Stokes $I$ profiles isolates the
Zeeman effect: the atomic lines broaden and, for the larger effective Land\'e
factors, split into resolved components, with strong antisymmetric $V$ and
symmetric $Q$ signatures of up to tens of percent of the continuum. Because the
field azimuth is zero, there is no direct contribution to Stokes $U$; the small $U$
signal arises entirely from Faraday rotation of the linear polarisation during
transfer through the atmosphere. The FeH bandhead is a contrasting molecular
case: its lines differ widely in magnetic sensitivity, so within a single segment
some are strongly broadened by the field while others are essentially unaffected,
and the net circular polarisation stays weak and complex as the many overlapping
components of opposite polarity cancel.

\begin{figure}[!th]
  \centering
  \includegraphics[width=\hsize]{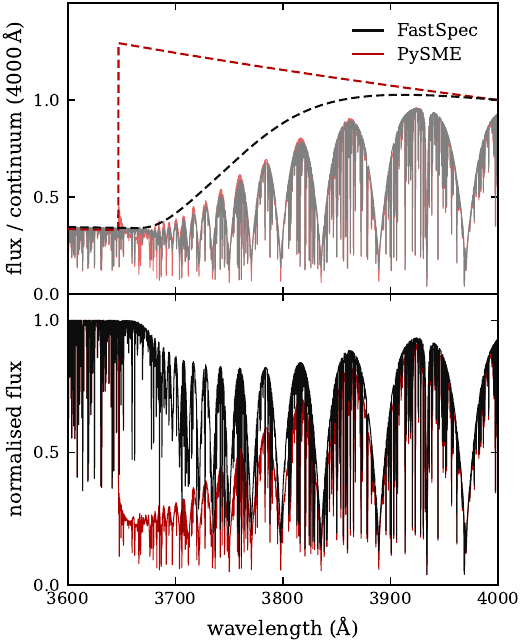}
  \caption{Balmer-jump region of an A-type model ($\Teff=9000$\,K,
    $\log g=4.0$) computed with \fastspec\ (grey) and \textsc{PySME} (red)
    for identical inputs. \emph{Top:} Flux
    (solid) and continuum (dashed) spectra of each code, divided by the respective
    continua at 4000\,\AA. \emph{Bottom:} Continuum-normalised spectra.
    }
  \label{fig:hlines}
\end{figure}

\paragraph{Hydrogen lines and the Balmer jump.}
Figure~\ref{fig:hlines} examines the Balmer-jump region of an A-type model
($\Teff=9000$\,K, $\log g=4.0$). We compare \fastspec\
with an independent synthesis produced by \pysme\
for the same model atmosphere, abundances, microturbulence, and atomic
line list. The spectra are shown for
the 3600--4000\,\AA\ interval that brackets the jump. The two
calculations differ only in their treatment of the hydrogen opacity:
\fastspec\ synthesises the Balmer lines with its occupation-probability
treatment of \citet{barklem:2003} and merges them smoothly into the bound-free continuum, whereas
\pysme\ superposes the individual high Balmer lines on a sharp hydrogenic
bound-free edge.

Shortward of the 3646\,\AA\ series limit and beyond $\simeq$\,4000\,\AA, the two codes agree to better than
$0.1\%$ in continuum-normalised flux, confirming that the metal-line and
lower-Balmer-line opacities are mutually consistent. They differ only over
3646--4000\,\AA, where \pysme\ produces a discontinuous Balmer jump at the
series limit whereas \fastspec\ yields a gradual one as the highest Balmer
lines dissolve into the continuum.

\section{Summary}
\label{sec:summary}

We have presented \fastspec, a fast LTE stellar spectrum synthesis code
that combines the physical completeness of the classical
model-atmosphere codes with the computational speed demanded by large
line lists and surveys, and with the polarised radiative transfer capability
required to model magnetic stars. Its physical ingredients -- the improved \sme\
equation of state, \atlas-family continuum opacities, the
occupation-probability hydrogen package, tabulated \ion{He}{i} Stark
broadening, and standard prescriptions for atomic and molecular line
broadening -- are drawn from well-tested public codes and restructured
for speed. The code owes its performance to four design choices:
comprehensive OpenMP parallelisation, adaptive per-line wavelength
windowing with weak-line rejection, memory-capped block processing of
arbitrarily large line lists, and table-based Voigt and Faraday-Voigt
profiles. \fastspec\ synthesises intensity and flux spectra, as well as
the full Stokes $IQUV$ vector when a magnetic field is present, and
handles atomic, molecular, and magnetic line lists in a single run. 
The code supports both plane-parallel and spherical model atmospheres, 
with either geometry compatible with full-Stokes polarised radiative transfer.
We have demonstrated the performance and range of applicability of \fastspec\ on the solar
spectrum, on the molecular bands of an M dwarf, on the polarised spectrum
of a magnetic star, and on the hydrogen lines of hot stars. The code is
portable, depends only on an HDF5-enabled Fortran compiler, and is
intended to serve both the modelling of individual stars and the
large-scale synthesis needs of surveys and spectropolarimetric 
magnetic field inversions.

\fastspec\ is accompanied by a suite of Python utilities that convert
\vald\ extractions and molecular line lists into its input formats and read
its binary output files into arrays for analysis and plotting. The code and these utilities
are made publicly available on GitHub.\footnote{\url{https://github.com/okochukhov/FastSpec}}

\begin{acknowledgements}
  This work made use of the \vald\ database, operated at Uppsala
  University and the University of Montpellier, and of the equation-of-state and hydrogen-opacity packages
  developed by P.\ Barklem, J.\ Valenti, and N.\ Piskunov. We thank B.\ Edvardsson
  for sharing his collection of molecular line lists and partition functions.
  We are grateful to the referee, C.\ Sneden, for a careful reading of the
  manuscript and constructive comments.
  Support by the Swedish Research Council (grant agreement no. 2023-03667) and the Swedish National Space Agency
  is acknowledged.
\end{acknowledgements}

\begin{appendix}

\onecolumn
\nolinenumbers

\section{Performance comparison with \textsc{Korg}}
\label{app:korg}

\textsc{Korg}\footnote{\url{https://github.com/ajwheeler/Korg.jl}}
\citep{wheeler:2023} is a modern open-source spectrum
synthesis code that, like \fastspec, is written for speed and supports
native multi-threaded execution, which makes it a natural reference point
for assessing the performance of \fastspec. The two codes do not share
the same physics genealogy -- they differ in the equation of state, the
continuum opacities, and the details of the line broadening -- so we make
no attempt at a point-by-point comparison of the emergent spectra.
Instead we configure both codes as closely as possible for three large
benchmark syntheses, matching the line lists, wavelength ranges, and
sampling, using the same \textsc{Marcs} model atmospheres, and
setting equivalent line-inclusion thresholds, and we compare the total
wall-clock time and the peak resident memory. Both codes were run on the
same machine (different from the one used for the timings in Table~\ref{tab:timing}). 
\textsc{Korg} (version 1.2.1) was launched with \texttt{julia -t auto},
which activates all logical cores, and \fastspec\ likewise used all OpenMP
threads. To reproduce the internal line-selection cutoff of \textsc{Korg},
we set the \fastspec\ line-to-continuum opacity threshold to
$\varepsilon=3\times10^{-4}$ (Sect.~\ref{sec:windowing}). Both codes read
their line lists from native HDF5 files -- \fastspec\ from its converted
\vald\ and molecular lists and \textsc{Korg} from lists saved in its own
HDF5 format -- so that the reported times reflect the synthesis rather than
the parsing of large ASCII line lists.

The first test is a solar synthesis over $3000$--$25\,000$\,\AA\ at a constant step $\Delta\lambda=0.01$\,\AA\
($2.2\times10^{6}$ wavelength points), with 
the full \vald\ atomic line list ($1.18\times10^{6}$ lines over the range). 
As summarised in Table~\ref{tab:korg},
\fastspec\ completes the synthesis in $12.9$\,s using $3.65$\,GB, against
$45.4$\,s and $14.05$\,GB for \textsc{Korg}, a factor of $3.5$ in speed
and $3.8$ in memory.
The second test targets the molecule-dominated spectrum of a cool M~dwarf
($\Teff=3500$\,K, $\logg=5.0$) over $5000$--$16\,000$\,\AA\ at $\Delta\lambda=0.01$\,\AA\ ($1.1\times10^{6}$
points), using the five TiO isotopologue line
lists ($^{46}$TiO to $^{50}$TiO) also employed in Sect.~\ref{sec:tests},
$1.2\times10^{7}$ lines in total. Here \fastspec\ runs in $13.1$\,s using
$4.67$\,GB, whereas \textsc{Korg} takes $49.1$\,s and $11.40$\,GB,
corresponding to factors of $3.7$ in speed and $2.4$ in memory.
The third test is the most computationally demanding and isolates
the molecular opacity that shapes the near-infrared spectra of cool
dwarfs. Using the same M-dwarf model, we synthesise the
$9500$--$25\,000$\,\AA\ region at $\Delta\lambda=0.01$\,\AA\
($1.55\times10^{6}$ points) with the \exomol\ water line list, which
contributes $1.31\times10^{8}$ lines over this interval. \fastspec\ completes the synthesis
in $66.4$\,s using $16.56$\,GB (close to the user-defined 16\,GB RAM limit applied in this test), 
against $649.2$\,s and $16.71$\,GB for
\textsc{Korg}, a factor of $9.8$ in speed at essentially the same peak
memory.

Across the three tests, \fastspec\ is between $3.5$ and $9.8$ times
faster than \textsc{Korg}, the advantage growing with the size of the
line list and reaching almost an order of magnitude for the water
synthesis, where line handling dominates the total cost. Its memory
footprint is $2$--$4$ times smaller in the atomic and TiO tests and
comparable to that of \textsc{Korg} in the H$_2$O test, where the shared
cost of storing more than $10^{8}$ lines dominates the working set of
both codes. These gains are
consistent with the
design choices described in Sect.~\ref{sec:performance}: adaptive line
windowing makes the cost scale with the number of significant
line-wavelength interactions rather than with the product of list size
and grid size, while memory-limited block processing keeps the working
set bounded. These results are not a criticism of \textsc{Korg}, whose
design emphasises automatic differentiability and interoperability within
the Julia ecosystem and therefore adopts different implementation
trade-offs; both codes are orders of magnitude faster than the legacy
synthesis programs they build upon.

\begin{table}[!t]
\caption{Timing and peak memory of \fastspec\ and \textsc{Korg} for three
benchmark syntheses.}
\label{tab:korg}
\centering
\small
\begin{tabular}{l c c c c c c c}
\hline\hline
 & & & & \multicolumn{2}{c}{$T_{\rm all}$ (s)} & \multicolumn{2}{c}{RAM (GB)} \\
\cline{5-6}\cline{7-8}
Test & Range (\AA) & $N_\lambda$ & $N_{\rm lines}$ & \fastspec & \textsc{Korg} & \fastspec & \textsc{Korg} \\
\hline
Sun / atomic  & 3000--25\,000 & $2.20\times10^{6}$ & $1.18\times10^{6}$ & \phantom{0}12.9 & \phantom{0}45.4 & \phantom{0}3.65 & 14.05 \\
M dwarf / TiO & 5000--16\,000 & $1.10\times10^{6}$ & $1.20\times10^{7}$ & \phantom{0}13.1 & \phantom{0}49.1 & \phantom{0}4.67 & 11.40 \\
M dwarf / H$_2$O & 9500--25\,000 & $1.55\times10^{6}$ & $1.31\times10^{8}$ & \phantom{0}66.4 & 649.2 & 16.56 & 16.71 \\
\hline
\end{tabular}
\tablefoot{Both codes were run on the same hardware 
using all logical cores. $N_\lambda$ is the
number of wavelength points, $N_{\rm lines}$ the number of lines in the
input list over the range, $T_{\rm all}$ the total wall-clock time, and RAM the peak memory footprint.}
\end{table}

\section{Input parameter file}
\label{app:input}

\begin{table*}[!t]
\caption{Input keywords for the \fastspec\ control file.}
\label{tab:keywords}
\small
\centering
\begin{tabular}{l p{12cm} l}
\hline\hline
Keyword & Meaning & Default \\
\hline
\texttt{WAVE}      & Start and end wavelength, step, and mode (0: constant
                     in wavelength, 1: constant in velocity) & -- \\
\texttt{SAVE}      & Five flags selecting the outputs: line-strength list,
                     line opacity, continuum opacity, intensity, flux & -- \\
\texttt{LINES}     & Block of line-list entries, one per line: species code
                     and file name & -- \\
\texttt{MODEL}     & Model-atmosphere file name (\texttt{.krz}) & -- \\[0.1cm]
\texttt{MUS}       & Number of $\mu$ angles, or an explicit list of $\mu$
                     values & 7 \\
\texttt{VMIC}      & Microturbulent velocity (\kms) & $0.0$ \\
\texttt{CONVOL}    & $\vsini$, $\vmac$ (\kms) and Gaussian width (\kms) or
                     resolving power & none \\
\texttt{BFIELD}    & Magnetic field vector (G): three numbers for \texttt{BFRAME} 0 or 1,
                     or $2N_\mu+1$ numbers for \texttt{BFRAME} 2 & none \\
\texttt{BFRAME}    & Field vector frame: stellar (0), observer (1), or a
                     list of inclination/azimuth pairs (2) & 0 \\
\texttt{[M/H]}     & Metallicity scaling of the default solar abundance pattern & -- \\
\texttt{ABUNDANCE} & Individual element abundances overriding those in the model header/default solar pattern & -- \\
\texttt{OPACITY}   & Precomputed line-opacity file to add & none \\
\texttt{LIMITRAM}  & RAM limit (GB) for the large arrays & $16$ \\
\texttt{EPSOPAC}   & Line-to-continuum threshold $\varepsilon$ for line
                     inclusion & $10^{-4}$ \\
\texttt{VWFUDGE}   & Scaling factor of the Uns\"old van der Waals width & $1.0$ \\
\texttt{H2BRD}     & Include H$_2$ in the van der Waals broadening & no \\
\texttt{MOLBRD}    & Apply Stark/van der Waals broadening to
                     molecular lines & no \\
\texttt{BFKURUCZ}  & Choice of hydrogen bound-free cross-sections
                     (\atlas\ or \textsc{Marcs}) & \atlas \\
\hline
\end{tabular}
\end{table*}

The input parameter file uses a \texttt{keyword=value} syntax; comment
lines beginning with \texttt{\#} or \texttt{!} and inline comments are
ignored. The recognised keywords are summarised in Table~\ref{tab:keywords}.
The mandatory keywords are \texttt{WAVE}, \texttt{SAVE}, \texttt{LINES},
and \texttt{MODEL}; the remaining keywords are
optional and take the default values given in the last column.

\section{Output files}
\label{app:output}

The outputs are written as unformatted Fortran stream files, each
beginning with the array dimensions. The line-strength list
(\texttt{fs\_linst.bin}) contains one block per line list, each holding the
species names, ionisation stages, central wavelengths, and central line
depths of the retained lines. The line- and continuum-opacity files
(\texttt{fs\_lineop.bin}, \texttt{fs\_contop.bin}) store the wavelength
grid, the opacity as a function of depth and wavelength, and the model
depth structure. The intensity file (\texttt{fs\_spec.bin}) stores the
$\mu$ grid, the wavelength grid, and the line and continuum intensities at
each angle -- or, in the magnetic case, the four Stokes parameters and the
continuum. The flux file (\texttt{fs\_flux.bin}) stores the wavelength grid
and the disk-integrated line and continuum fluxes. Python routines that
read these files into arrays are included in the distribution.

\end{appendix}

\end{document}